\documentclass[aps,prl,twocolumn,superscriptaddress,showpacs]{revtex4}

\usepackage{graphicx}
\usepackage{multirow}
\usepackage[usenames]{color}

\begin{document}

\title{Breakdown of thermodynamic equilibrium for DNA hybridization
in microarrays}

\author{J. Hooyberghs}
\affiliation{Flemish Institute for Technological Research (VITO),
Boeretang 200, B-2400 Mol, Belgium}
\affiliation{Institute for Theoretical Physics,
KULeuven, Celestijnenlaan 200D, B-3001 Leuven}
\affiliation{Hasselt University, Campus Diepenbeek,
B-3590 Diepenbeek, Belgium}
\author{M. Baiesi}
\author{A. Ferrantini}
\author{E. Carlon}
\affiliation{Institute for Theoretical Physics,
KULeuven, Celestijnenlaan 200D, B-3001 Leuven}

\date{\today}

\begin{abstract}
Test experiments of hybridization in DNA microarrays show
systematic deviations from the equilibrium isotherms. 
We argue that these deviations are due to the presence of a partially
hybridized long-lived state, which we include in a kinetic model.
Experiments confirm the model predictions for the intensity vs. free
energy behavior.  The existence of slow relaxation phenomena has important
consequences for the specificity of microarrays as devices for the
detection of a target sequence from a complex mixture of nucleic acids.
\end{abstract}


\maketitle

\begin{table}[b]
\caption{Target and probe sequences used in the experiments. All sequences
have a 5' to 3' orientation. Target sequences have a 20-mer poly(A) stretch
attached to their 3' end, which terminates with a Cy3 fluorophore. Probes
have a 30-mer poly(A) stretch at their 3' end, which is covalently linked 
to the microarray surface.} 
\begin{tabular}{lr}
\hline
 \multicolumn{2}{l}{Target sequences in solution}\\
\hline
1. &CTTTGTCGAGCTGGTATTTGGAGAACACGT \\
2. &{TCGAGCTGGTATTTGGAGAACACGT}\\
\hline
 \multicolumn{2}{l}{Probes at the microarray surface}\\
\hline
PM & ACGTGTTCTCCAAATACCAGCTCGACAAAG\\
\multirow{4}{*}{1MM}&ACGTG\underline{A}TCTCCAAATACCAGCTCGACAAAG\\
&ACGTG\underline{C}TCTCCAAATACCAGCTCGACAAAG\\
&ACGTG\underline{G}TCTCCAAATACCAGCTCGACAAAG\\
& \multicolumn{1}{c}{\ldots} \\
\multirow{2}{*}{2MM}&ACGTG\underline{A}TCTCC\underline{C}AATACCAGCTCGACAAAG\\
& \multicolumn{1}{c}{\ldots} \\
\hline
\end{tabular}
\label{table1}
\end{table}

DNA hybridization (the binding of two strands to form a double helix)
in bulk solution has been extensively studied in the past \cite{bloo00}.
In this letter we discuss hybridization in DNA microarrays. 
\textcolor{Black}{Microarrays
are high throughput devices which have been widely used to measure the
activity of genes at a genome wide level. In a microarray singled stranded
DNAs are arrayed on a solid substrate in spots, each containing
a specific sequence.}
Hybridization takes place between surface-bound sequences
(referred to as {\it probes}) and sequences in solution ({\it
targets}) carrying a fluorophore.  The amount of hybridized target is
obtained from the emitted fluorescence from a given spot.
Although hybridization in microarrays has attracted
some interest in recent years its physical properties are still poorly
understood \textcolor{Black}{(for reviews on the topic see, e.g.,
\cite{reviews}).} We demonstrate here that, contrary to a widespread
belief, in DNA microarrays equilibration times may largely exceed typical
experimental times.  These claims are based \textcolor{Black}{on
experimental results and are corroborated by the analysis of a kinetic
model.}

The experimental setup is shown in Table~\ref{table1} and extends that of
Ref.~\cite{hooy09_prl}. A single sequence is present in solution: either
a 30-mer or a 25-mer. The surface probe sequences
are perfect matching, with one or two mismatches. The mismatches can
be of different nature and they are at different positions along the
sequence \cite{last_note}.
In total there are 1006 different probe
sequences \cite{hooy09_prl}. Custom arrays containing spots with the probe
sequences of Table~\ref{table1} were purchased from Agilent Technologies.
\textcolor{Black}{We used 15K slides which accommodate 15 replicas of
the 1006 sequences. The analysis is performed on the median intensities
over the replicas. The standard Agilent protocol (except for target
fragmentation) and Agilent buffers \cite{agi} were used. The temperature
is $65^\circ$~C.}

\begin{figure}[t]
\includegraphics[width=7.0cm]{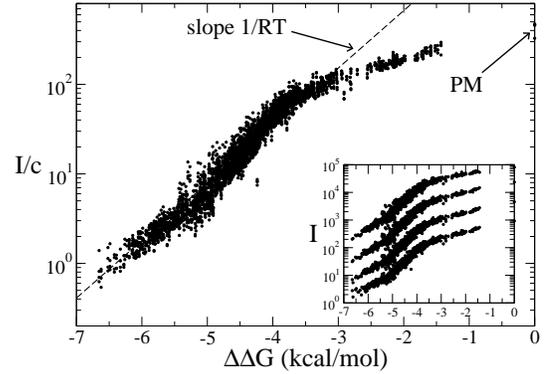}
\caption{Plot of $I/c$ as a function of $\Delta \Delta G$ for four
different experiments at concentrations $c=2$, $10$, $50$ and $250$
pM. The hybridization time is of 17 h.  \textcolor{Black}{The
``collapse" of the four data sets into a single curve demonstrates that $I
\propto c$ in the whole intensity range.} Deviations from the equilibrium
isotherm $I \propto e^{-\Delta G/RT}$ are observed at high intensities.
Inset: plot of each individual concentration.
}
\label{fig_IvsDG_andc}
\end{figure}

Figure~\ref{fig_IvsDG_andc} shows a plot of $I/c$
(\textcolor{Black}{the intensity divided by the target concentration})
vs.~$\Delta \Delta G$ for four experiments at different concentrations
using the setup of Table~\ref{table1} (for the 30-mer target). The
variable $\Delta
\Delta G \equiv \Delta G_{PM} - \Delta G$ is the difference in
hybridization free energies between a given sequence and the perfect
match (PM) sequence.  It is calculated from the nearest-neighbor
parameters obtained from the analysis of microarray data, as discussed
in Ref.~\cite{hooy09_prl} (the nearest-neighbor model assumes that
$\Delta G$ can be written as a sum of dinucleotide terms \cite{bloo00}).
\textcolor{Black} 
{Figure~\ref{fig_IvsDG_andc} shows that the intensity
is proportional to the concentration for four orders of magnitude in $I$.
In the low $c$ limit, equilibrium thermodynamics predicts that
\begin{equation}
I = A c e^{-\Delta G/RT}
\label{langmuir}
\end{equation}
where $A$ sets the intensity scale, $R$ is the gas constant and $T$ the
temperature. Eq.~(\ref{langmuir}) is obtained from the $c\to 0$ limit
of the Langmuir isotherm (which was used in microarray data analysis
\cite{held03_prl,naef03,bhan03_sh,carl06,nais09}), but also from isotherms
in which electrostatic effects are taken into account \cite{vain02}
(electrostatic effects were experimentally observed at low ionic strengths
\cite{gong08}). In the latter $\Delta G$ contains a contribution from
electrostatic interactions.  For both isotherms, in the $I \propto c$
regime one expects a linear dependence of $\log I$ on $\Delta G$ (or
$\Delta \Delta G$).  Fig.~\ref{fig_IvsDG_andc} shows that the experimental
data are only in partial agreement with Eq.~(\ref{langmuir}), which is
drawn as a dashed line in the figure.  
}

\begin{figure}[t]
\includegraphics[width=8.0cm]{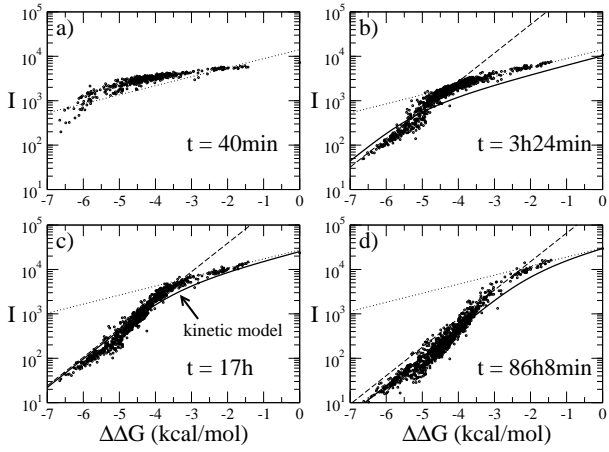}
\vspace*{2mm}
\caption{Plot of intensity vs. $\Delta \Delta G$ four experiments at
different times with a $30$-mer target and $c=50$ pM. Dashed lines have
slope $1/RT$, 
\textcolor{Black}{dotted}
lines have slope $\gamma/RT$ with $\gamma = 0.32$.
Solid lines are obtained from the solution of Eqs.~(\ref{dtheta1dt}) and
(\ref{dtheta2dt}) using 
\textcolor{Black}{as parameters $k_1 =10^5 M^{-1} s^{-1}$ ,
$k_2=1s^{-1}$, $\gamma = 0.32$ and $\Delta G_{\rm PM} = -14.5$ 
kcal/mol}.
}
\label{fig_timesL30}
\end{figure}

\begin{figure}[t]
\hspace*{-8mm}
\vspace*{2mm}
\includegraphics[width=8.0cm]{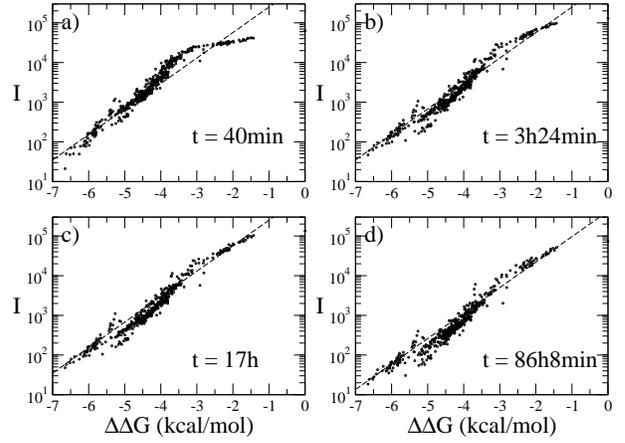}
\caption{As in Fig.~\ref{fig_timesL30} for a target sequence of
length $25$. 
\textcolor{Black}{The target concentration is $c=500$pM}.}
\label{fig_timesL25}
\end{figure}

We then extended the analysis at different hybridization times.
Figure~\ref{fig_timesL30} shows a plot of $I$ vs. $\Delta \Delta G$
for a $30$-mer target at four different times and for a concentration
of $50$ pM (the $17$~h hybridization data are those already shown in
Fig.~\ref{fig_IvsDG_andc}). Once the desired hybridization time has been
reached the experiment is stopped, the microarray washed and scanned
to measure the emitted fluorescence from every spot.  Experiments at
different hybridization times thus require different slides.  As the
hybridization time increases, a larger fraction of the data aligns along
a line with a slope $1/RT$, which shows that the observed deviations from
Eq.~(\ref{langmuir}) are due to the breakdown of thermodynamic equilibrium.
Surprisingly, full equilibrium has not been reached here even after
$86$h.  \textcolor{Black}{In Fig.~\ref{fig_timesL30}(b,c,d) the
data for $\log I$ align along two slopes: in the equilibrium regime the
slope is $1/RT$ (Eq.~(\ref{langmuir})), in the non-equilibrium regime
the slope is smaller and appears to be constant in the course of time
\cite{note_slope}.} Hybridization data for the shorter target sequence
(25-mer) are shown in Fig.~\ref{fig_timesL25}.  The agreement with
Eq.~(\ref{langmuir}) is over three orders of magnitude in the intensity
scale at times $> 3$h.
\textcolor{Black}
{Hence equilibration is much faster for the shorter
sequences. The experimental setup allows a detection of non-equilibrium
effects as deviation from the $1/RT$ line without the need, in principle,
of a time series analysis.  
}

\begin{figure}[t]
\includegraphics[width=6cm]{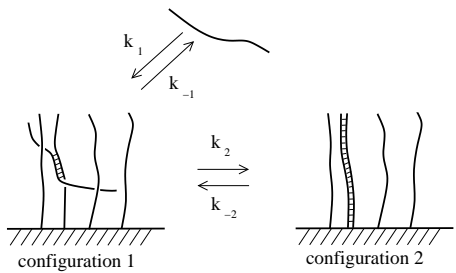}
\caption{The three state model for hybridization in DNA
microarrays is specified by the four rate constants.}
\label{fig3state}
\end{figure}

Hybridization 
of oligonucleotides in solution is usually described as a two state process.
However, as will be shown, a two-state process cannot be reconciled with the experimental data shown in Fig.~\ref{fig_timesL30}.
During manufacturing the
probes are tethered to the surface 
and can form a dense
\textcolor{Black}{layer} that slows down
hybridization. \textcolor{Black}{Some of these effects have been
discussed in Refs.\cite{pete01,haga04_prl}}.  The typical
distance between probes is $10$ nm, and the length of a
fully stretched 30-mer duplex is $10$ nm and its thickness of $2$
nm. Probe sequences in the experiment have also a poly(A) 30-mer spacer
(see Table~\ref{table1}). Therefore a single target molecule can interact
with more than one probe. Taking this into account, we have extended
the two state hybridization model with an additional intermediate state
(Fig.~\ref{fig3state}).  Indicating with $\theta_1$ and $\theta_2$,
the fraction of partially and fully hybridized probes on a microarray
spot, the kinetics of these reactions is given by
\begin{eqnarray}
\frac{d\theta_{1}}{dt} &=& c k_1 (1-\theta_1 -\theta_2 ) +
k_{-2} \theta_{2} - (k_{-1} + k_2) \theta_1
\label{dtheta1dt}\\
\frac{d\theta_2}{dt} &=& k_2 \theta_1 - k_{-2} \theta_2
\label{dtheta2dt}
\end{eqnarray}
where $c$ is the target concentration in solution and $k_1$, $k_{-1}$,
$k_2$ and $k_{-2}$ the four rates involved (see Fig.~\ref{fig3state}). For
simplicity we have assumed that at most a single target molecule can
bind to a given probe.
\textcolor{Black}{
$\theta_1$ is the average occupation fraction over several configurations
in which partial binding can occur at different positions of the probe
sequence. 
}

The rate constants, using a two state model description, have been
measured in several microarray experiments. 
In Ref.~\cite{glaz06_sh}
the hybridization of a common target sequence to a perfect match probe and
to a probe containing one mismatch were considered. The following rates
were measured (at $45^\circ$~C): $k_1^{\rm (PM)} = 19 \cdot 10^{4} M^{-1}
s^{-1}$, $k_1^{\rm (MM)} = 21 \cdot 10^{4} M^{-1} s^{-1}$, $k_{-1}^{\rm
(PM)} = 12 \cdot 10^{-4} s^{-1}$ and $k_{-1}^{\rm (MM)} = 29 \cdot 10^{-4}
s^{-1}$. While there is more than a factor two of difference in the
detachment rates, the attachment rates differ only by $10\%$.
These results are in agreement with observation for kinetic behavior
in bulk solution \cite{cant80}. The probes in our experiment differ by
at most two nucleotides out of 30.
\textcolor{Black}{We will then take $k_1$ as sequence independent.
Consider now the reaction rate $k_2$. In the partially hybridized state
the target strand binds over a stretch of nucleotides with one probe
sequence (primary contact) but it can also bind to a second neighboring
probe (secondary contact).  The rate limiting step is the unbinding from
secondary contacts and the strand contraction so that the target probe
can overcome steric hindrance and wind up into a fully formed helix all
along its length. $k_2$ will depend on target length and probe length
and density.
We will assume $k_2$ to be the same for the probes of Table~\ref{table1}.
}
The reverse rates are then fixed by the thermodynamics relations
\begin{equation}
k_{-1} = k_1 e^{\Delta G'/RT}, \ \ \ \ \ \
k_{-2} = k_2 e^{(\Delta G - \Delta G')/RT}
\label{k1km1}
\end{equation}
where $\Delta G'$ and $\Delta G$ are the free energy differences
between configurations 1 and 2, and the unbound state, respectively.
\textcolor{Black}{
Next we link $\Delta G'$ to $\Delta G$.
Weak total binding (small $|\Delta G|$) caused by the presence of
multiple mismatches should also correspond to weak partial binding
(small $|\Delta G'|$). As a simple approximation we will assume that 
the two free energies are monotonically linked as 
}
\begin{equation}
\Delta G' \approx \gamma \Delta G \ \ \ \ \ \ \ \  (\gamma < 1) \,.
\label{intro_gamma}
\end{equation}
The model is thus characterized by $k_1$, $k_2$ and $\gamma$.


To gain some more insight we consider the limit of fast equilibration
for Eq.~(\ref{dtheta1dt}). First we obtain the equilibrium value for
$\theta_1$ by setting the right hand sides of Eqs.~(\ref{dtheta1dt})
and (\ref{dtheta2dt}) to zero 
\textcolor{Black}{in}
the limit $c e^{-\Delta G/RT} \ll 1$.
We then solve Eq.~(\ref{dtheta2dt}) replacing for $\theta_1$ its equilibrium 
value \textcolor{Black}{$\theta_1^{\rm (eq)} =c e^{-\Delta G'/RT}$}. 
Setting the initial condition $\theta_2 (0)=0$ we get
\begin{equation}
\theta_2 (t) =  c e^{-\Delta G/RT} \left( 1 -  e^{-t/\tau} \right),
\label{solution}
\end{equation}
\begin{eqnarray}
\tau^{-1} = k_{-2} = k_2 e^{(\Delta G - \Delta G')/RT} =
k_2 e^{(1 - \gamma)\Delta G/RT}.
\label{eq:tau}
\end{eqnarray}
The relaxation time, $\tau$, depends on $\Delta G$: weakly bounded
sequences (small $|\Delta G|$) equilibrate faster than strongly
bounded ones (large $|\Delta G|$). For fast equilibrating sequences
($\tau \ll t$) one recovers \textcolor{Black}{Eq.~(\ref{langmuir})}
from Eq.~(\ref{solution}); for sequences with long equilibration times
$\tau \gg t$ we expand Eq.~(\ref{solution}) to lowest order in $t/\tau$.
With this approximation we find that for a given time $t$
\begin{eqnarray}
\theta_2 (t) =
\left\{
\begin{array}{ccc}
c e^{-\Delta G/RT} & & |\Delta G| \ll |\Delta G^*| \\
c t k_2 e^{-\gamma \Delta G/RT} & & |\Delta G| \gg |\Delta G^*|
\end{array}
\right.
\label{fastk1}
\end{eqnarray}
where $\Delta G^*$ is a crossover free energy that depends on
time and is obtained by setting $\tau = t$ in Eq.~(\ref{eq:tau}).
\textcolor{Black}{After hybridization the slides undergo 
washing steps according to the standard Agilent protocol, which
are expected to remove weakly bound target
molecules from the slide and have been also included in thermodynamics
models of arrays \cite{wash}. In the present setup there is only one
target sequence in solution and washing is likely to affect the partial
hybridized state. In this case one can assume that the measured intensity
is given by $I \approx A \theta_2$ (in the model the typical free
energies of the partially hybridized states are such that 
$\theta_1 \ll \theta_2$).
}
Equation~(\ref{fastk1}) reproduces the two slopes in the
$\log I$ vs. $\Delta \Delta G$ plots as seen in the experiments
(Fig.~\ref{fig_IvsDG_andc}). It shows that the non-equilibrium regime
is characterized by a slope equal to $\gamma/RT$.

The solid lines in Fig.~\ref{fig_timesL30}(b,c,d)
are plots of the intensity $I = A \theta_2$ obtained
from the solution of Eqs.~(\ref{dtheta1dt}) and (\ref{dtheta2dt}).
\textcolor{Black}{The
parameters used are given in the caption of Fig.~\ref{fig_timesL30}.}
For the choice of parameters given, 
the fast equilibration limit (Eq.(\ref{solution})) approximates 
very well the full solution of Eqs.~(\ref{dtheta1dt}) and (\ref{dtheta2dt}).
There is a reasonable agreement between the experiments and the kinetic
model. We note though that the crossover between the two regimes is
somewhat sharper in experiments. In addition, the experimental $I$
vs. $\Delta \Delta G$ data show a slight sigmoidal trend which is not
present in the kinetic model.
\textcolor{Black}{Within
the two state model kinetics, and using the assumption $k_1$ as sequence
independent, one arrives to a solution similar to Eq.~(\ref{fastk1})
although with $\gamma=0$. Therefore the two state model cannot 
account for a second
finite slope as observed in the experiments.  }
\textcolor{Black}{
Note that the fit of the kinetic model to the data requires
also an estimate of $\Delta G_{\rm PM}$, as the method of
Ref.~\cite{hooy09_prl} does not provide absolute $\Delta G$ for a sequence
but only $\Delta \Delta G$, i.e. differences in free energies with respect
to a perfect match hybridization.  In addition, in
the fit we adjusted the constant
$A$ ($I = A \theta_2$) as we note in the experimental data a global
decrease of the intensity scale.  This is probably due to some degradation
of the fluorophores or of the target and probe strands (as depurination,
the loss of purines, which alters the binding energies of the involved
strands). The overall decrese in intensity occurs both for the 25-mer
and the 30-mer sequences. In the latter the effect is somewhat stronger,
especially at longer hybridization times.} 

\textcolor{Black}{
Eq.~(\ref{eq:tau}) predicts that the relaxation time is a function of
$\Delta G$, $\Delta G'$ and $k_2$.  We can use this equation to compare
the ratio between the times for a $L=30$ and $L=25$ targets. Consider
the same probe sequence hybridizing to the two targets.  Assuming $\Delta
G'(L=30) \approx \Delta G'(L=25)$ and using as estimate of the difference
in binding energies between $\Delta G(L=30)  - \Delta G(L=25) \approx
-2.5$ kcal/mol \cite{note1}, we get from Eq.~(\ref{eq:tau}) a decrease
of a factor $\exp(2.5/RT) \approx 40$ in the relaxation time. In addition
one also expects $k_2 (L=25) > k_2 (L=30)$ which decreases the relaxation
time even further. The similarities between Fig.~\ref{fig_timesL30}(c)
and Fig.~\ref{fig_timesL25}(a) suggest that the relaxation time ratio
between $25$-mers and $30$-mers is approximately $25$, which is the
same order of magnitude just obtained from Eq.~(\ref{eq:tau}). Since
typical biological experiments involve target strands of lengths 30-50,
the breakdown of equilibrium shown here may occur
in many different microarrays platforms and in biological experiments,
and could involve even longer relaxation times.  }

Summarizing: We showed that hybridization in DNA microarrays under
standard conditions is characterized by relaxation times which may
largely exceed the experimental time. In the non equilibrium regime the
intensities are distributed as $e^{-\gamma \Delta G/RT}$,
with $\gamma < 1$.  This is equivalent to introducing an effective
temperature $T_{\rm eff} = T/\gamma > T$. Interestingly effective
temperatures were used as adjustable phenomenological parameters to
fit biological microarray data \cite{held03_prl,carl06}. This work
provides an insight on the origin of these.

The breakdown of equilibrium implies lower specificity of the microarrays
as devices for the detection of a desired sequence from a complex
mixture. To see this consider a probe at the microarray surface and two
sequences  at equal concentration in solution: one perfect matching with
the probe and one with a mismatch. In the equilibrium regime the two
sequences hybridize to the probe with a probability ratio $ e^{(\Delta
G_{PM} - \Delta G_{MM})/RT} \approx 0.05$, where we have used a typical
value $\Delta G_{MM} - \Delta G_{PM} \approx 2$~kcal/mol \cite{hooy09_prl}
and a temperature of $T=65^\circ$~C. In the nonequilibrium regime, due
to the presence of a factor $\gamma<1$ in the exponential
the ratio is about $0.4$ (taking $\gamma=0.32$).  
Therefore in the non-equilibrium regime a
significant fraction of a measured signal may be due to hybridization to
non-complementary targets, a phenomenon known as cross-hybridization.
For an optimal functioning of the microarrays it is then desirable to
work under equilibrium conditions \cite{bhan03_sh}. Several parameters
may influence the relaxation time as temperature, salt and buffer
conditions. The experimental setup discussed in this paper provides a
good test of equilibrium (single line vs. broken line in a $I$ vs. $\Delta
\Delta G$ plot) and can be used to investigate the best working conditions
for hybridization.
\begin{acknowledgments}
We thank Karen Hollanders (VITO) for help with the experiments.
We acknowledge financial support from Research Fundation -- Flanders
(FWO) G.0311.08 and from KULeuven OT/07/034A.
\end{acknowledgments}


\begin{thebibliography}{22}
\bibitem{last_note}
Up to 5 nucleotides to the edges to avoid terminal mismatches, and such
that the minimal distance between mismatches in the same strand is of
5 nucleotides.
\bibitem{note_slope}
\textcolor{Black}{
We note that the slope in the non-equilibrium regime of
Fig.~\ref{fig_IvsDG_andc}(a) is somewhat smaller than that of (b,c,d).
This is probably due to the protocol followed: at time $t=0$ both the
solution containing the target molecules and the array are at room
temperature. They are then placed into an oven for the duration of the
experiment at a temperature $T=65^\circ$~C.  Short time behavior
could then be influence by the initial room temperature hybridization.
}
\bibitem{note1}
\textcolor{Black}{
We estimated $\Delta G_{\rm PM}(L=30) = -14.5$kcal/mol from the fits of
Fig.~\ref{fig_timesL30}, using extensivity we expect $\Delta G_{\rm PM}(L=25) \approx
0.83 \Delta G_{\rm PM}(L=30) = -12$ kcal/mol
}
\end{thebibliography}

\end{document}